\documentclass[12pt]{iopart}

\usepackage{iopams}
\usepackage{graphicx}
\usepackage{epstopdf}
\usepackage{here}

\begin{document}

\title[Immersed transient eddy current flow metering]{Immersed transient 
eddy current flow metering: a
calibration-free velocity measurement technique 
for liquid metals}

\author{N Krauter and F Stefani}

\address{Helmholtz-Zentrum Dresden - Rossendorf, Bautzner Landstr. 400,
D-01328 Dresden, Germany}
\ead{F.Stefani@hzdr.de}
\begin{abstract}

Eddy current flow meters (ECFM) are widely used for
measuring the flow velocity of electrically 
conducting fluids. Since the flow
induced perturbations of a magnetic field depend
both on the geometry  and the 
conductivity of the fluid, extensive 
calibration is needed to get accurate results. 
Transient eddy current flow metering (TECFM) 
has been developed to overcome this problem.
It relies on tracking the 
position of an impressed eddy current system which is moving 
with the same velocity as the conductive fluid. 
We present an immersed version of this measurement 
technique and demonstrate its viability by
numerical simulations and a first experimental 
validation.

\end{abstract}

\vspace{2pc}
\submitto{\MST}
\maketitle

\section{Introduction}

Measuring the flow velocity of liquid metals is a 
challenging task because 
of their opacity, chemical reactivity and - in most cases - 
elevated ambient temperature \cite{ECKERT}. Fortunately, 
the high electrical conductivity of liquid metals often allows to 
use magnetic inductive measurement techniques. These 
techniques generally rely on applying magnetic fields to the 
fluid and measuring appropriate features, e.g. 
amplitudes, phases, or forces, of the flow induced magnetic 
fields.

A local embodiment of this technique is the eddy current 
flow meter (ECFM) as patented by Lehde and Lang in 1948
\cite{LELA}, which consists of
two primary coils excited by an AC generator, and 
one secondary coil located midway between them.
Modifications of this method, using one primary coil
and two secondary coils,  were described in 
\cite{SURESH,POORNA}.
Another version of this local sensor, which measures the
flow induced change of the amplitude in the vicinity
of a small permanent magnet, is the 
magnetic-distortion probe described by 
Miralles et al. \cite{MIRALLES}.

A global embodiment of the same principle, 
the contactless
inductive flow tomography (CIFT), is able to 
reconstruct entire
two or three-dimensional flow fields  
from induced field amplitudes that are measured 
at many position around the fluid when it is 
exposed to one or a few external magnetic fields 
\cite{MST1,CIFT,MST2}. 

Another inductive measurement concept relies on the 
determination of magnetic phase shifts due to the flow 
\cite{PRIEDE,BUCHENAU}. Further, the Lorentz force 
velocimetry (LFV) 
determines the force acting on a permanent magnet
close to the flow, which results as a direct consequence
of Newton's third law applied to the braking
force acting by the magnet on the flow \cite{THESS}.
With this technique, it is even possible to measure
velocities of fluids with remarkably low conductivities,
such as salt water \cite{HALBEDEL}.

A common drawback of (nearly) all those methods is 
that they require extensive calibration
since the flow induced magnetic field perturbations depend
both on geometric details of the measuring system and 
on the conductivity of the fluid, which is, in turn, 
temperature-dependent.
Actually, the signals are proportional to the magnetic
Reynolds number $Rm=\mu_0 \sigma V L$, where
$\mu_0$ is the magnetic permeability constant, $\sigma$ the
conductivity of the liquid, and $V$ and $L$ denote 
typical velocity and length scales of the relevant 
fluid volume.
Further to this, the use of permanent magnets, 
as necessary for the magnetic distortion probe \cite{MIRALLES}
and for LFV \cite{THESS}, or of magnetic yoke materials, 
as for the phase-shift method \cite{PRIEDE,BUCHENAU},
set serious limitations to the ambient temperature 
at the position of the respective sensors.

Transient eddy current 
flow metering (TECFM) \cite{FORBRIGER2} 
aims at overcoming both drawbacks. 
Building upon earlier work of Zheigur and 
Sermons \cite{ZHEIGUR}, 
this is accomplished by impressing a traceable 
eddy current system into the liquid metal and 
detecting its movement with appropriately 
positioned magnetic field sensors. 
Since the eddy current 
moves with the velocity of the liquid, there is no 
need for a calibration of the sensor.
The non-invasive TECFM sensor for measuring the liquid 
metal velocity close to the fluid boundary from outside,
as described in \cite{FORBRIGER2}, represents
a specific external realization of TECFM.  

Here, we present a  modified variant of TECFM, an invasive 
sensor that can be placed within a liquid metal pool or a 
pipe to measure the local velocity in the surrounding 
metal. After describing the main functioning principle of 
this {\it immersed transient eddy current flow metering} 
(ITECFM), we 
will illustrate the method by numerical simulations.
Then, first flow measurements in the eutectic alloy GaInSn 
will be presented. The paper closes with 
some conclusions and a discussion 
of the prospects to use the method under high temperature 
conditions.

\section{The principle of ITECFM}

ITECFM is intended to measure the local velocity or the flow rate 
around the sensor in liquid metal pools or large pipes (for 
small pipes there will be some distortion of the results, when the 
penetration depth of the magnetic field into the liquid metal 
is larger than the radius of the pipe). Basically, the ITECFM sensor 
is an invasive tube-shape sensor which is put inside the liquid metal, 
parallel to the flow direction. There is no direct contact 
between the liquid metal and the pick-up coils because 
the latter are protected by a cladding, made of stainless 
steel for example. In contrast to the external variant of TECFM
\cite{FORBRIGER2}, this configuration traces the zero crossing of 
the magnetic field of the eddy current system instead of the 
position of a magnetic pole. For this purpose, the coils are 
arranged differently in order to allow a velocity measurement 
of the surrounding liquid.

\begin{figure}[H]
	\centering
	\includegraphics[width=14cm]{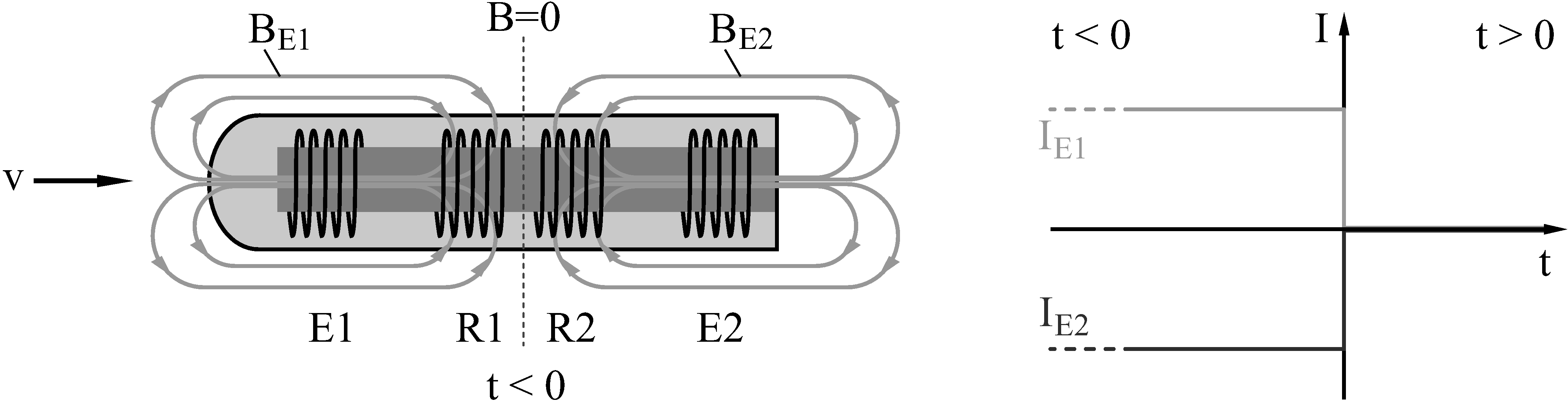}
	\caption{Basic structure of the ITECFM sensor and 
	its magnetic field produced by the two 
	excitation coils E1, E2 before the excitation 
	currents are switched off. The induced voltages 
	are measured by the two receiver coils R1 and R2.
	The time dependent excitation currents $I_\mathrm{E1}$ and 
	$I_\mathrm{E2}$ are displayed on the right hand side.}
	\label{Fig:image1}
\end{figure}

\begin{figure}[H]
	\centering
	\includegraphics[width=14cm]{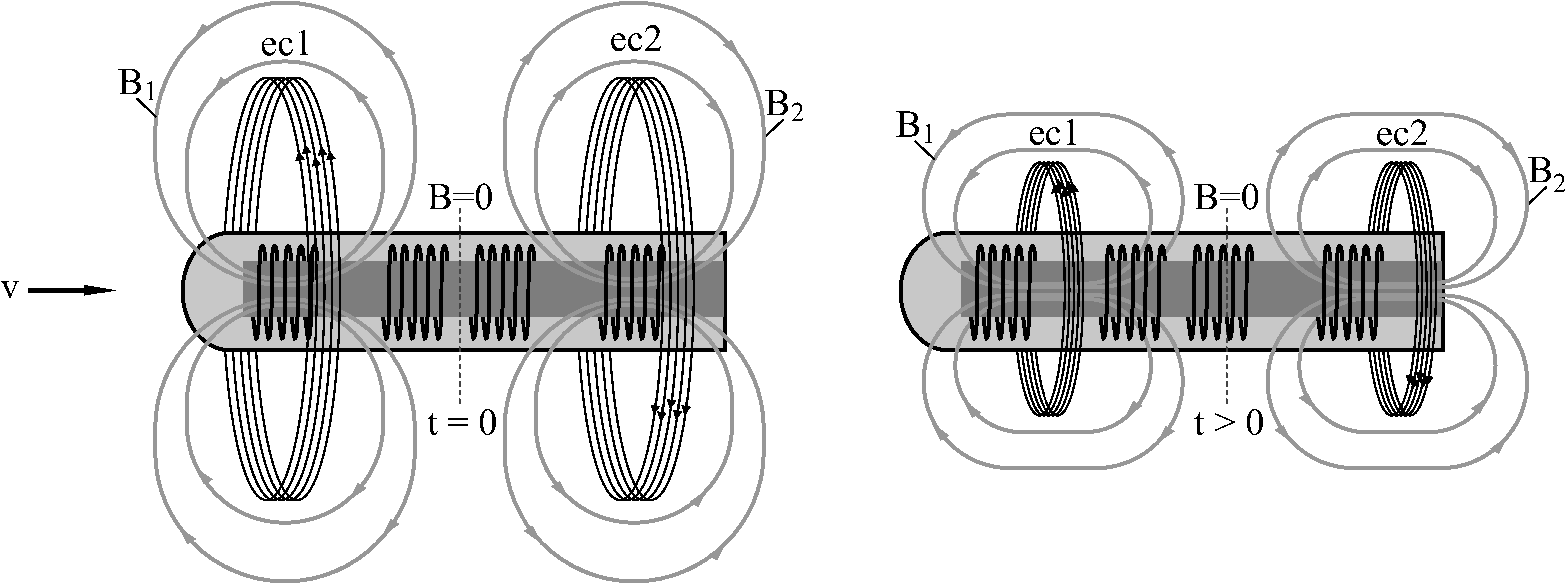}
	\caption{Immersed TEC-FM: Eddy currents ec1 and ec2 
	induced within the liquid metal and their magnetic 
	fields $B_\mathrm{1}$, $B_\mathrm{2}$ when the 
	excitation currents are switched off at $t=0$ (left) 
	and for $t>0$ (right) when the eddy currents have 
	started to move in flow direction and begin to dissipate.}
	\label{Fig:image1a}
\end{figure}

The eddy currents within the liquid metal, which are to be 
used for inferring the fluid velocity, are induced by the 
excitation coils E1 and E2 (see figure~\ref{Fig:image1}). 
Figure \ref{Fig:image1a} shows a simplified scheme of these eddy currents. Both magnetic 
fields $B_\mathrm{E1}$ and $B_\mathrm{E2}$ are generated by 
current steps which occur at the same time, but in opposite 
directions. The result are two oppositely directed  
magnetic fields with 
the same amplitude, which will induce opposing eddy currents 
during switching on or off of the excitation currents. Because 
of the symmetric arrangement of the coils (and, therefore, 
the magnetic fields), 
the zero crossing $x_\mathrm{0}$ of the total magnetic field 
$B$ is located exactly in the middle between the receiver 
coils R1 and R2, when $v_\mathrm{liquid}$ is zero and/or immediately 
after the current step for $v_\mathrm{liquid}>0$. 
The excitation currents are assumed to be 
switched off at $t=0$.

Although the eddy currents ec1, ec2 and their magnetic fields 
$B_\mathrm{1}$, $B_\mathrm{2}$ are dissipating \cite{FORBRIGER1}, 
for $v_\mathrm{liquid}=0$
the zero 
crossing remains exactly in their middle, regardless of 
their magnitude. This changes, however, if the fluid is moving. Then, the 
eddy currents are transported in flow 
direction with the velocity of the fluid. Under the reasonable 
assumption that the electrical conductivity of the liquid metal 
is homogeneous around the sensor, both eddy currents will 
dissipate with the same rate and the zero crossing of the magnetic 
field will also move with the fluid velocity. The position of the 
zero crossing can be tracked by means of 
the receiver coils R1 and R2.

Just as in \cite{FORBRIGER2}, the position of the zero 
crossing can be calculated according to
\begin{eqnarray}
\label{f1}
x_0(t)&=&\frac{x_1 \dot{B}_\mathrm{2}(t) - x_2\dot{B}_\mathrm{1}(t)}{\dot{B}_\mathrm{2}(t)-\dot{B}_\mathrm{1}(t)}
=\frac{x_1 U_2(t) - x_2U_1(t)}{U_2(t)-U_1(t)}
\end{eqnarray}
where $x_1$ and $x_2$ are the positions of the receiver 
coils R1 and R2, and $U_1$ and $U_2$ are the 
respective voltages measured there. Although the arrangement of the coils 
for the external TECFM is different, this simplified 
formula can be used to approximate the liquid metal 
velocity in the case of ITECFM, too. This will be validated by 
numerical simulations in the next section.

\section{Numerical simulations}

The simulations of ITECFM were implemented in COMSOL 
Multiphysics 5.0, using a time dependent 2D axisymmetric model 
and the magnetic fields (mf) physics environment.

\subsection{Simulation Model}

For the simulation, some simplifications have been made. 
The flow velocity $v_\mathrm{liquid}$ of the liquid metal is 
assumed constant and homogeneous around the sensor thimble.  
The liquid metal does not contain foreign particles or gas 
bubbles. Furthermore, any Lorentz forces exerted by the 
excitation coils on the liquid metal are neglected.

\begin{figure}[H]
	\centering
	\includegraphics[width=11cm]{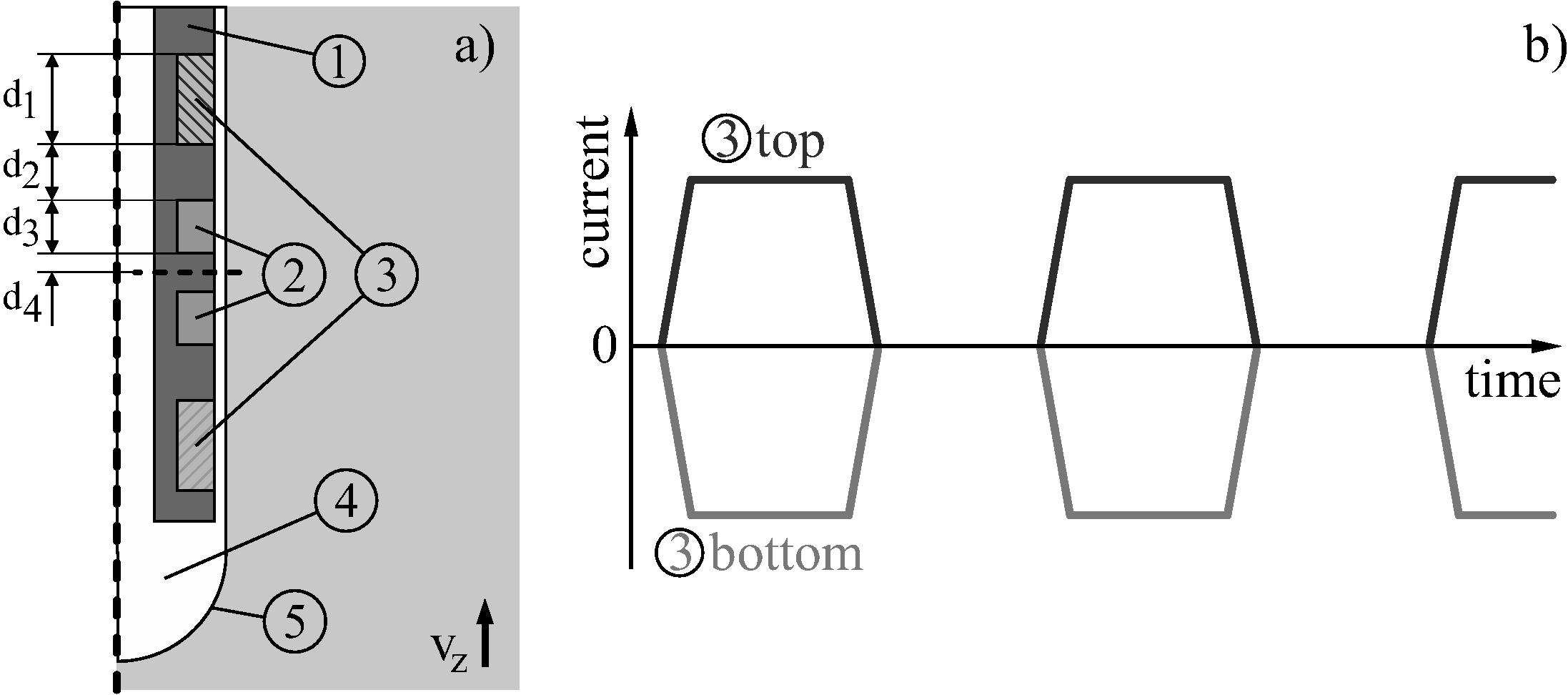}
	\caption{a) Simplified simulation model of the 
	ITECFM sensor: 1 - coil holder (core), 2 - receiver coils, 
	3 - excitation coils, 4 - air, 5 - sensor thimble, 
	$v_z$ indicates the flow direction of the liquid 
	metal, the dashed vertical line represents the symmetry 
	axis for the 2D axisymmetric model, the dashed horizontal 
	line shows the position of the centre of symmetry 
	regarding the coils ($d_\mathrm{1}=5$\,mm, 
	$d_\mathrm{2}=3$\,mm, $d_\mathrm{3}=3$\, mm, 
	$d_\mathrm{4}=1$\,mm). b) Excitation current pulses 
	for the top and bottom excitation coil}
	\label{P:SimModel}
\end{figure}

For an optimal operation of the sensor, the receiver 
coils should 
be arranged symmetrically, with the centre of symmetry 
exactly in the middle between the two  excitation coils 
(see dashed horizontal line in figure \ref{P:SimModel}). 
Reasonable positions of the receiver and excitation coils 
have been determined by multiple simulations with 
variations in arrangement, size and spacing between the 
coils. In principle, there are two possibilities for the 
arrangement of the coils: the receiver coils can be placed 
between the excitation coils or vice versa. However, since 
the initial position $x_\mathrm{0}$ of the zero crossing 
of the magnetic field is located exactly in the middle 
between the excitation coils, the receiver coils should be 
placed as close as possible to this point in order to 
achieve maximum sensitivity of the sensor. Placing the
excitation coils between the receiver coils is also possible but 
would result in a much lower signal amplitude and sensitivity 
because of the increased distance from $x_\mathrm{0}$.

\begin{figure}[H]
	\centering
	\includegraphics[width=9cm]{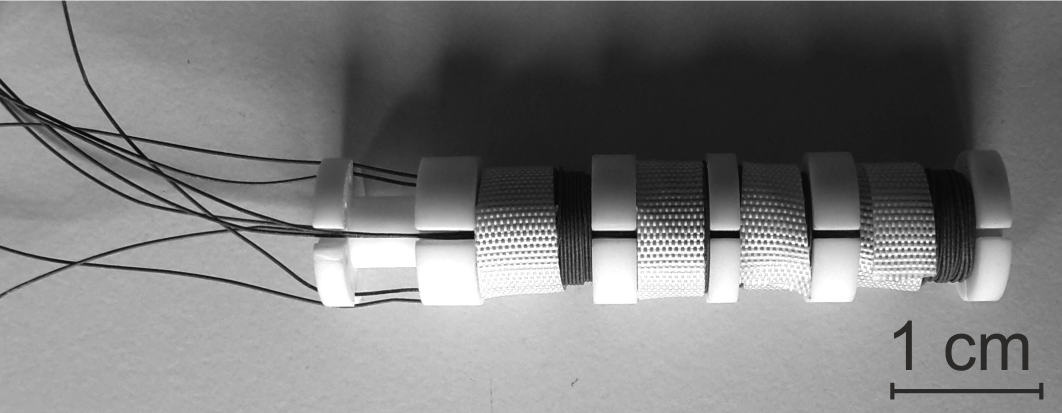}
	\caption{High temperature prototype of the ITECFM sensor 
	for temperatures up to 650~$^{\circ}$C.}
	\label{Fig:foto}
\end{figure}

The size of the excitation coils turn out to 
have only a minor influence 
on the functionality of the sensor, 
as long as the absolute values of 
the excitation current pulses are the same, since 
$x_\mathrm{0}$ is always in the 
middle between them. 
Their actual size was chosen to accommodate a 
reasonable number of turns for the coil wires in the actual 
prototype. However, the axial extension of the 
receiver coils should be as 
small as possible because this will increase their 
sensitivity for detecting the zero crossing of the 
magnetic field, and also minimize the dependence on the 
conductivity of the liquid.
Further to this, since the distance between the coils and the 
boundary of the liquid metal has a significant influence on the signal 
strength, the air gap between coils and inner wall 
of the sensor, as well as the wall thickness of the sensor 
thimble, should be as small as possible.
The actual sizes, turn numbers, and wire thicknesses 
of the coils for a low temperature (LT) 
and a high temperature (HT)
prototype  of the ITECFM sensor 
are shown in table 1 (see also 
figure \ref{P:SimModel} a). A photography of the high 
temperature sensor is shown in figure~\ref{Fig:foto}.

\begin{table}[H]
	\centering
	\caption{Properties of the low temperature(LT) and high temperature(HT) prototypes }
	\begin{tabular}{|l|c|c|}
		\hline  & LT Sensor & HT Sensor  \\
		\hline
		\hline Receiver coils: height & 3~mm & 5~mm  \\
		\hline Excitation coils: height & 5~mm & 8~mm  \\
		\hline
		\hline Receiver coils: turns & 120 & 80  \\
		\hline Excitation coils: turns & 100 & 120  \\
		\hline
		\hline Receiver coils: wire thickness & 0.15~mm & 0.25~mm  \\
		\hline Excitation coils: wire thickness & 0.25~mm & 0.25~mm  \\
		\hline
		\hline Max. operation temperature & $150^{\circ}$\,C & $650~^{\circ}C$  \\
		\hline Core material & PVC & ceramic  \\
		\hline Wire material & copper & 73\% copper, 27\% nickel  \\
		\hline Wire isolation & enamel & ceramic  \\
		\hline
	\end{tabular}
\end{table}

\subsection{Simulation results}

The edge steepness of the excitation current plays an important role 
for ITECFM because the fluid 
velocity can only be extracted from the magnetic fields when the 
excitation currents have reached their final value, i.e. 
$dI/dt = 0$. 
From this instant on, R1 and R2 detect exclusively the magnetic fields 
of the eddy currents within the liquid. 

\begin{figure}[H]
	\centering
	\includegraphics[width=9cm]{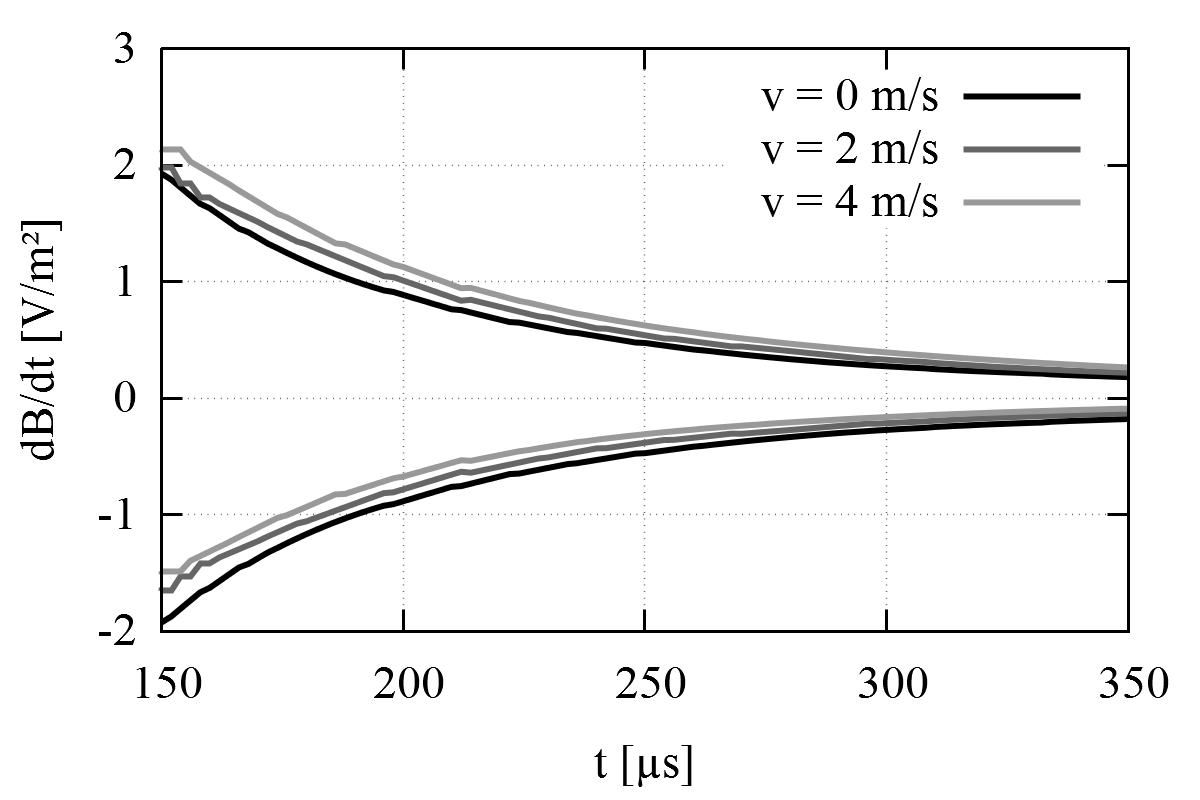}
	\caption{Simulation results of $\dot{B}_\mathrm{1}$ 
	and $\dot{B}_\mathrm{2}$ as a function of time, computed 
	for three fluid velocities 
	$v_\mathrm{liquid}$ and the conductivity of 
	GaInSn $\sigma=3.3 \times 10^6$\,S/m. 
	The excitation current is supposed to be 
	switched off at $t=0~\mu$\,s and to reach zero at 
	$t=100~\mu$\,s.}
	\label{Fig:image2}
\end{figure}

In figure \ref{Fig:image2} we see how the time derivatives 
of the fields $B_1$ and $B_2$ at the two receiver coils R1 and R2 
change with increasing fluid velocity. For 
this example, the three top curves show the values for 
$\dot{B}_1$ and the three bottom curves the values 
for $\dot{B}_2$. While they are symmetrical for 
$v_\mathrm{liquid}=0$, 
there is a growing asymmetry between $\dot{B}_1$ and $\dot{B}_2$ for
increasing $v_\mathrm{liquid}$.      

\begin{figure}[H]
	\centering
	\includegraphics[width=14cm]{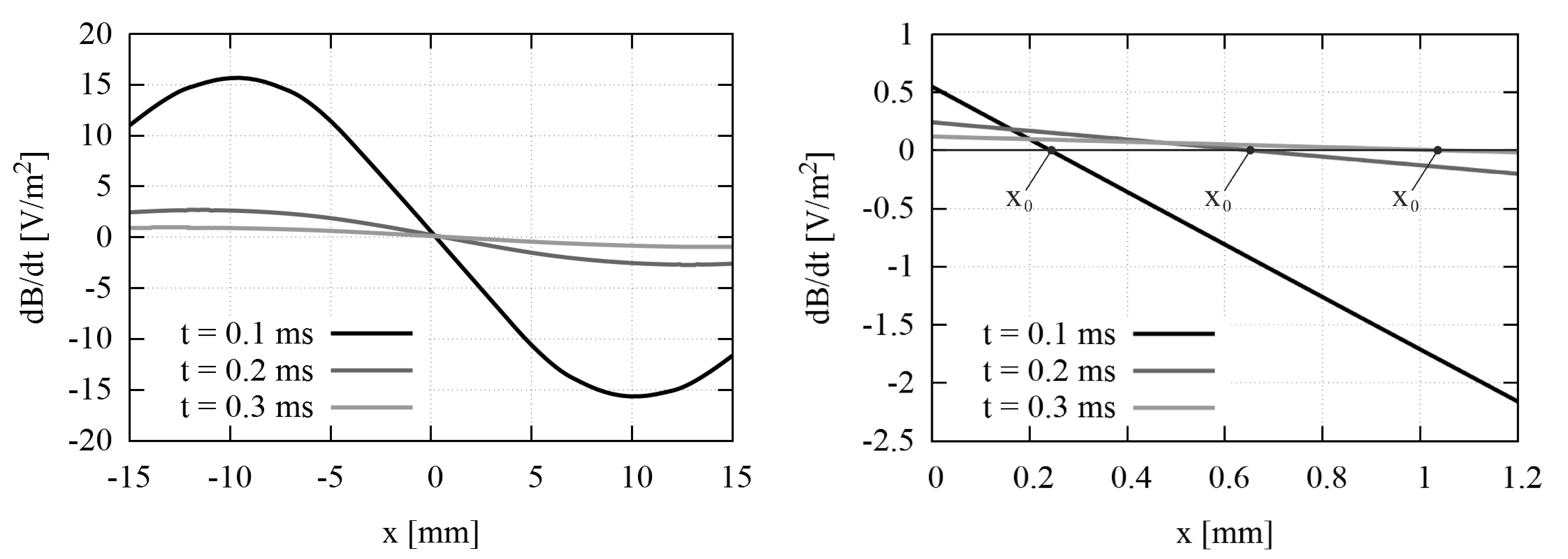}
	\caption{Simulation results of the time derivative of the 
	total magnetic field for $v_\mathrm{liquid}=4$\,m/s 
	in a range of $\pm 15$\,mm around the centre of the 
	sensor (left), and a zoom to $x=0$\,mm (right) 
	after 0.1~ms, 0.2~ms and 0.3~ms. $x=0$\,mm is located 
	exactly in the middle between the coils R1 and R2. 
	$x_\mathrm{0}$ shows the time-dependent 
	position of the zero crossing 
	of $\dot{B}$.}
	\label{Fig:image8}
\end{figure}

When plotting $\dot{B}$ on a line parallel to the flow direction 
at different instants in time after the current 
steps, the movement 
of the zero crossing can clearly be seen (figure \ref{Fig:image8}). 
Although the magnetic field is significantly 
dissipating with time, 
the zero crossing of $\dot{B}$ keeps moving with 
$v_\mathrm{liquid}$. This is shown in the right panel of 
figure \ref{Fig:image8} , 
where $x_0$ marks the time-dependent position of the zero crossing. 

\begin{figure}[H]
	\centering
	\includegraphics[width=9cm]{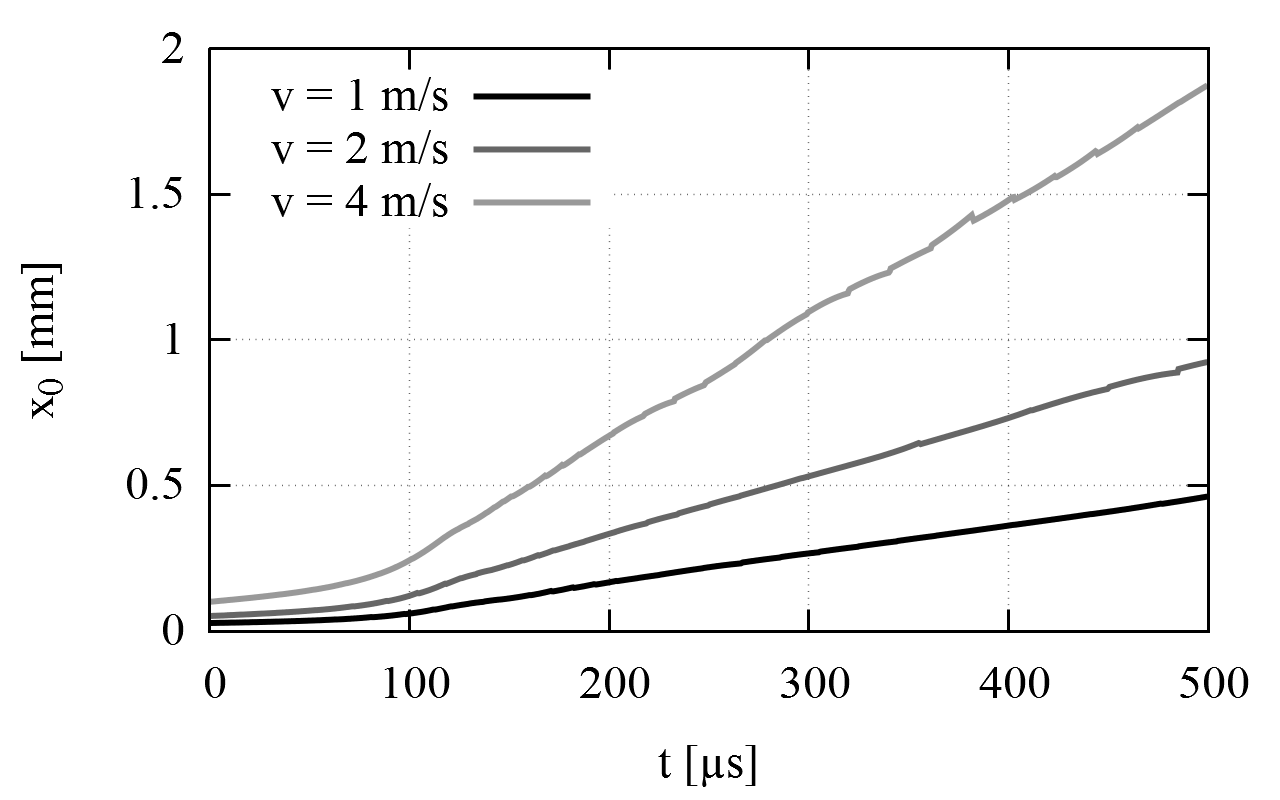}
	\caption{Simulation results for the time dependent 
	position $x_\mathrm{0}(t)$ for three different velocities. 
	The current steps reach their final value after $100~\mu$s.}
	\label{Fig:image9}
\end{figure}

Figure \ref{Fig:image9} shows the movement 
of $x_\mathrm{0}$ over time. 
The fluid velocity is then inferred from 
the slope of the respective 
line in the data set. Since, in this particular 
simulation,  the excitation currents 
reach zero only at $100~\mu$\,s, before that instant
$x_\mathrm{0}(t)$ appears to move slower than the 
fluid velocity.

The reason for this effect is 
the superimposition with the remaining 
 magnetic fields of the 
excitation coils. 
For laboratory experiments it is therefore 
advisable to use a current source 
with a high edge steepness. Otherwise 
$B_1$ and $B_2$ would have already dissipated too 
much for an accurate measurement.

While we have considered the case of switching off the
excitation current, almost the same results are 
obtained for switching on the currents. Although the magnetic 
fields of the excitation coils are not zero, they are constant after 
they reach their final value and would not induce currents within 
the excitation coils or the liquid metal. Yet, the difference
between both methods  would become larger for increasing $Rm$. 

\begin{figure}[H]
	\centering
	\includegraphics[width=14cm]{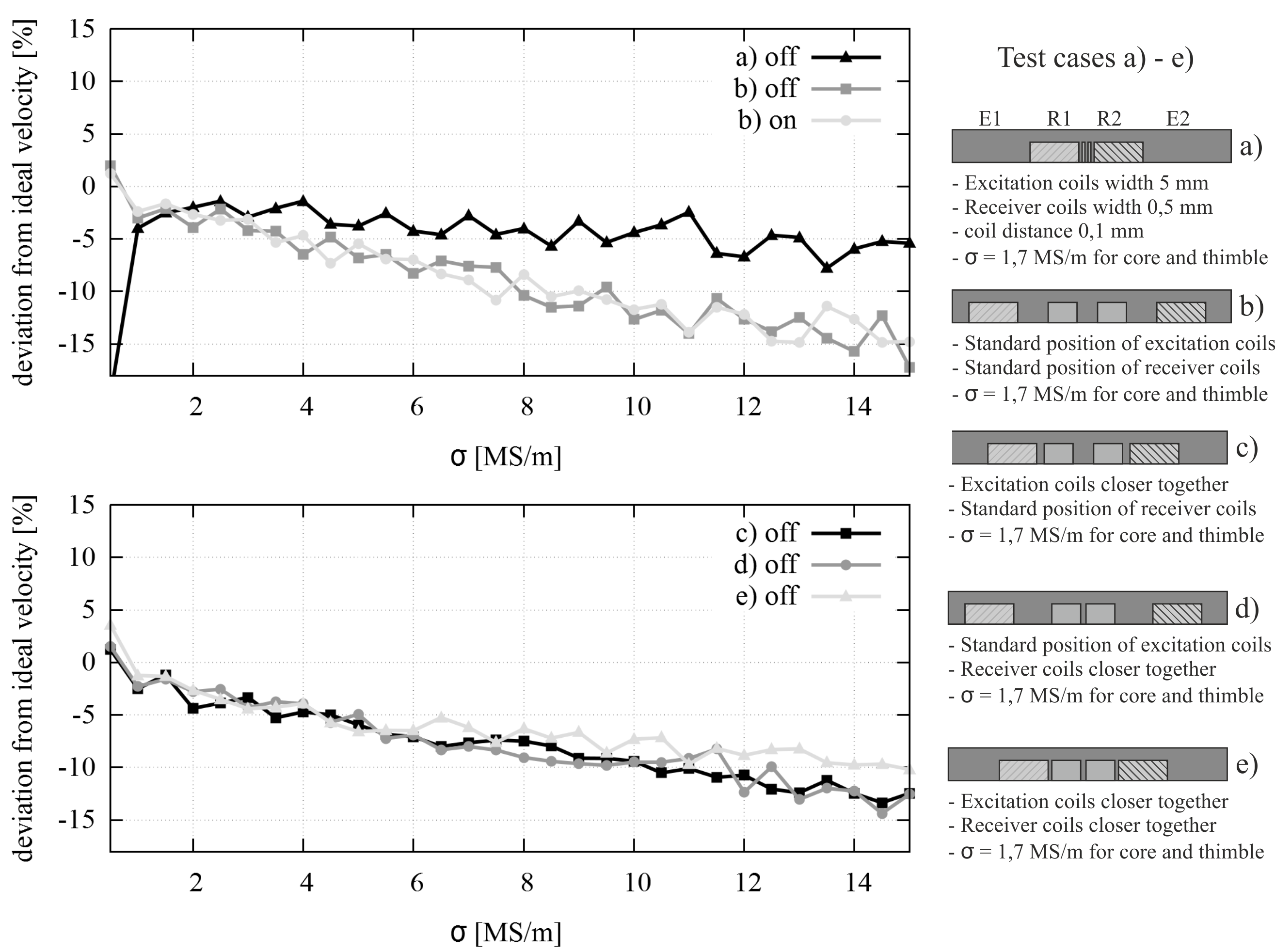}
	\caption{Simulation results for the deviation of the 
	measured velocity from the pre-given velocity for
	different coil geometries and varying 
	electrical conductivity of the liquid metal for 
	$v_\mathrm{liquid}=2$\,m/s when switching off (on) 
	the excitation current.}
	\label{Fig:image10}
\end{figure}

Until now, the simulations have been calculated only for one 
electrical conductivity of the liquid metal 
($\sigma=3.3 \times 10^6$\,S/m for GaInSn). 
With view on the strong influence
of the (temperature dependent) conductivity 
for conventional ECFM's, 
we present in figure \ref{Fig:image10} 
the calculated velocity for different 
conductivities
and a variety of hypothetical (and real) 
coil geometries.

In general, despite some slight dependence on the electrical conductivity of 
the fluid, the deviation from the ideal velocity is relatively weak
for a wide range of conductivities. Hence, ITECFM can essentially 
be considered as calibration-free. In the case of extremely thin 
receiver coils (a), the 
overall deviation would remain less than 5\% for 
$1$\,MS/m $< \sigma < 10$\,MS/m.
For the standard geometry (as embodied in the 
prototypes for which the coil size and distance 
between the coils have to be larger in order to 
accommodate a suitable number of turns and to 
facilitate the construction of the core) it can 
be seen (b and c) that the results for switching on 
or switching 
off the excitation currents are almost the same. 
The further simulation (c,d,e) show also that the 
deviation from the ideal 
velocity is smaller, when the coils are positioned as 
close to each other as possible, especially at high 
conductivities. At low conductivities of the liquid metal, 
the positioning of the coils has only a small influence 
on the results. The increasing deviations for higher 
fluid conductivities are related to the size of and 
the distance between the coils as well as the dissipation 
of the eddy currents. 

\begin{figure}[H]
	\centering
	\includegraphics[width=10cm]{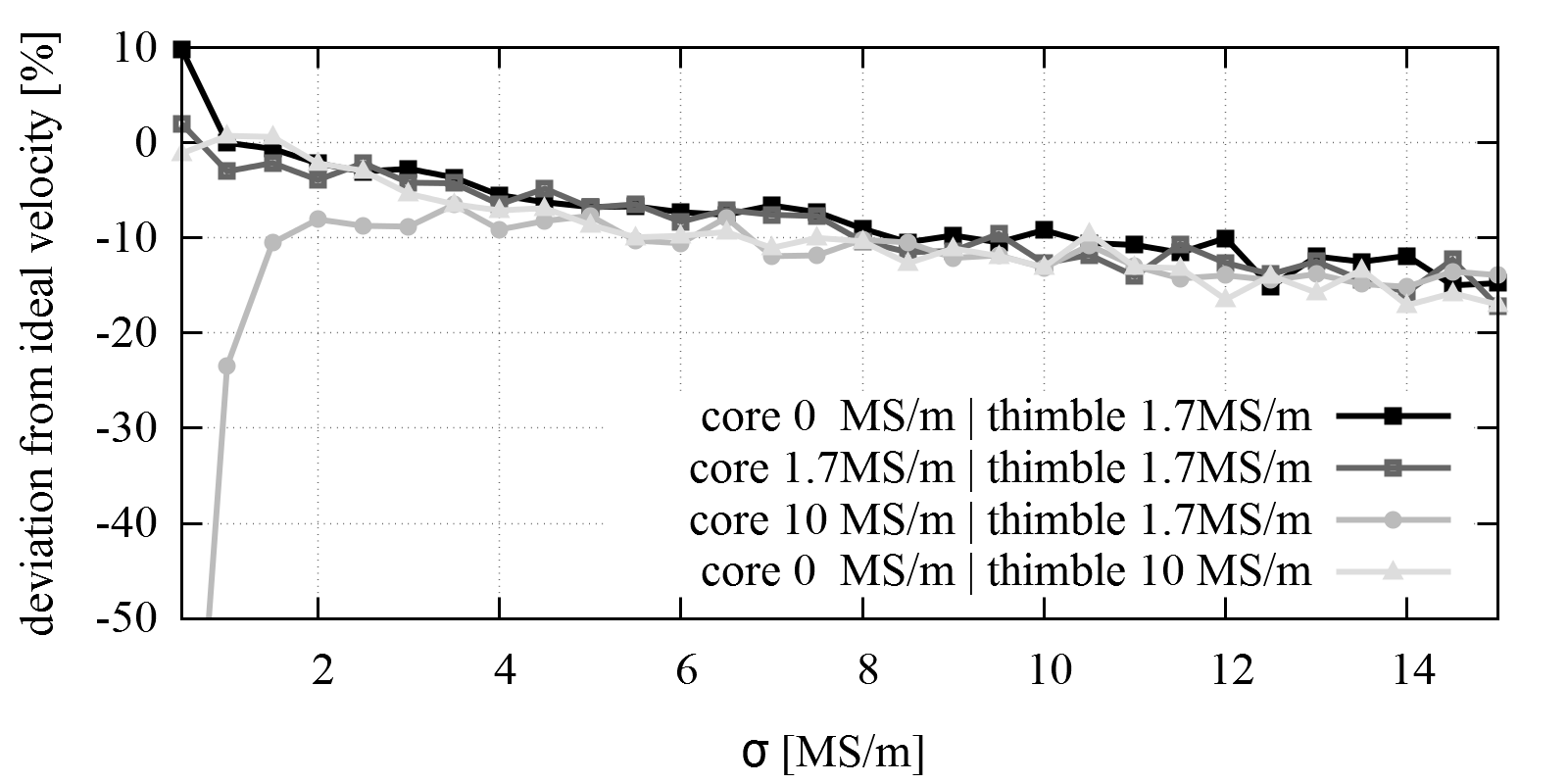}
	\caption{Simulation results for the deviation of the 
	measured velocity from the pre-given velocity for
	different electrical 
	conductivities of sensor core and thimble and varying 
	electrical conductivity of the liquid metal for 
	$v_\mathrm{liquid}=2$\,m/s when switching off 
	the excitation current. The coils 
	are arranged like in the standard case b).}
	\label{corewall}
\end{figure}

As can be seen in figure \ref{corewall}, the electrical 
conductivity of the sensor core, which holds the coils, 
and of the sensor thimble have a significant influence on 
the measurement results for the velocity, especially at 
low $\sigma$ of the liquid metal. The eddy currents which 
are induced within the conductive components of the sensor, 
are stronger for higher electrical conductivities. Unlike the 
eddy currents within the liquid metal, they are not moving 
with $v_\mathrm{liquid}$ but are stationary at all times. Because 
the magnetic fields of the eddy currents from liquid metal are
superimposed with the fields 
of the sensor components, the measured velocity appears to be 
lower. This effect is stronger at low $\sigma$ of the liquid 
metal because the eddy currents within the sensor components 
are in the same order of magnitude or even larger than the 
ones in the liquid metal. Another aspect to consider is the 
volume of the respective sensor components. Because the core 
has a considerable larger volume than the thimble wall, its 
conductivity has a larger influence on the velocity measurement. 
At higher $\sigma$ of the liquid metal the effect gets more 
and more negligible because of the stronger eddy currents 
and the larger volume of the liquid metal.

\section{Experimental results}

A first test of a  ITECFM sensor was carried out with the 
low-temperature prototype in a 
liquid metal loop with the eutectic alloy GaInSn. 
This sensor has a plastic coil holder, 
the excitation coils have 100 turns, the receiver coils have 
120 turns, and conventional copper wire of diameter 0.25~mm 
was used (see table 1). Rectangular 
voltage pulses of $5~V$ with a frequency of 1~kHz, a duty cycle 
of 50~\% and a fall time of $20~\mu$s have been used to generate 
the excitation currents.

\begin{figure}[H]
	\centering
	\includegraphics[width=11cm]{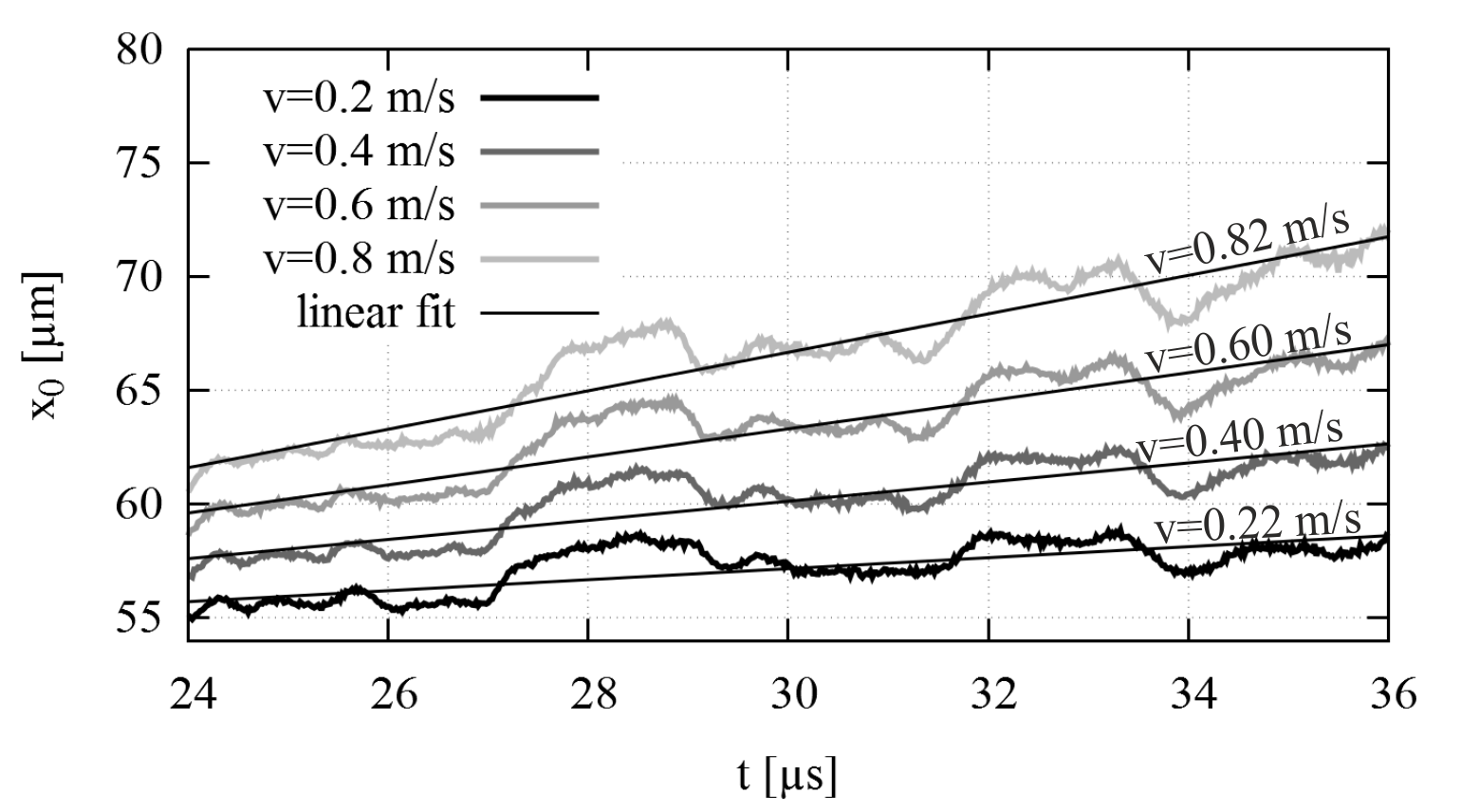}
	\caption{Measurement results for the movement of $x_\mathrm{0}$ for 
	ITECFM in a pipe flow of GaInSn. The fluid 
	velocity is inferred from the slope of the respective linear 
	fit for each pre-adjusted velocity. The inferred velocity 
	is indicated above each curve.}
	\label{Fig:image3}
\end{figure}

The sensor was put inside a stainless steel tube to prevent direct 
contact with the liquid metal. The receiver voltages were measured 
with a memory oscilloscope and $x_\mathrm{0}(t)$ was calculated with 
equation (\ref{f1}). Figure \ref{Fig:image3} shows the measurement 
results for four different fluid velocities and a linear fit of 
each dataset. The displayed results represent 
the mean value of 2500 measurement 
sweeps with one measurement taken every millisecond for $2.5s$. 
As can be seen in the previous section in figure 
\ref{Fig:image9}, the results for $x_\mathrm{0}(t)$ are expected 
to have a linear rise.

There are some deviations from the expected results for $x_\mathrm{0}(t)$, 
especially the disturbances around $t=28~\mu$s and $t=33~\mu$s. The overall 
slope of the linear fit however, is very close to the pre-adjusted 
flow velocity in the GaInSn-Loop (which is, as a matter of fact, also
not exactly known). The disturbances appear at the same times for each 
measurement and are most likely caused by the resonant frequency of the 
receiver coils.  Future 
experiments using tailored current sources instead of the
presently used voltage source are expected to 
improve this situation.


\section{Conclusions and prospects}
In this paper, we have presented the principle of 
ITECFM and some promising results obtained both 
in simulations as well as in an experiment 
with a first prototype in GaInSn. 
While its calibration-free 
character makes the method a promising candidate for a 
number of laboratory and industrial applications, it 
certainly needs further tests and optimization. Although 
both the external and immersed configurations of TECFM 
are based on the same 
principle, there are  some differences with view on the 
different arrangement of the excitation and receiver coils, 
which have to be addressed in detail. 

Future work will be devoted to more experiments with optimized 
excitation schemes and different 
liquid metals to validate the simulation 
results and the calibration free-free character of the sensor. 
Another advantage of ITECFM is the avoidance of 
any magnetic materials which makes it particularly suited for
high temperature applications. Tests with the high temperature 
prototype consisting of heat resistant materials 
are planned for ambient temperatures of up to 
650~$^{\circ}$C as they are typical, e.g., 
for sodium fast reactors.

\ack
This work was supported by CEA in the framework of the 
ARDECo programme.

\end{document}